\documentclass[aps,prb,twocolumn,showpacs,preprintnumbers,amsmath,amssymb,superscriptaddress]{revtex4}

\usepackage{graphicx}
\usepackage{dcolumn}
\usepackage{bm}

\begin{document}

\title{Orbital Kondo effect in double quantum dots}

\author{Piotr Trocha}
\email{ptrocha@amu.edu.pl}\affiliation{Department of Physics,
Adam Mickiewicz University, 61-614 Pozna\'n, Poland}

\date{\today}

\begin{abstract}
Orbital Kondo effect in a system of two single-level quantum dots
attached to external electron reservoirs is considered
theoretically. The dots are coupled {\it via} direct hoping term
and Coulomb interaction. The Kondo temperature is evaluated from
the scaling approach and slave boson technique. The later method
is also used to calculate linear conductance of the system.
Nonlinear conductance, in turn,  is calculated in terms of the
nonequilibrium Green function formalism.

\end{abstract}

\maketitle

\section{Introduction}{\label{Sec:1}}

Kondo effect in electronic transport through quantum dots (QDs)
strongly coupled to external leads is a many body phenomenon which
has been extensively studied in the last two
decades\cite{cronewett,gores,glazman,meir2,kang,meir3,aguado,aono,swirkowicz1,kuzmenko,lim}.
Spin fluctuations in the dot, generated by coupling of the dot to
external leads, give rise to a narrow peak in the dot's density of
states (DOS) at the Fermi level. This Rado-Suhl resonance results
in enhanced transmission through the dot, and leads to the unitary
limit of the linear conductance $G$ at zero temperature,
$G=2e^2/h$. The enhanced transmission is suppressed when a bias
voltage is applied, and this leads to the so-called zero-bias
(Kondo) anomaly in differential conductance. The above described
phenomenon arises from the two-fold spin degeneracy, and is often
referred to as the spin Kondo effect. However, the Kondo
phenomenon may also appear when the spin degree of freedom is
replaced by any two-valued quantum number, e.g. the one associated
with an orbital degree of freedom (the orbital Kondo
effect)\cite{zawadowski}. A minimal realization of the orbital
(spinless) Kondo phenomenon requires two orbital discrete levels
coupled to external leads\cite{boese,imry}. This can be realized
for instance in two single-level quantum dots coupled to external
electrodes\cite{wilhelm,hubel,sunG,sztenkiel,sztenkiel2,holle,krychowski,shon,wen,kubo}.
Coherent superposition of virtual tunneling events, in which one
electron tunnels from the dot QD1 (QD2) to one of the leads and
then simultaneously another electron tunnels to the dot QD2 (QD1),
leads to the Kondo resonance at low temperatures.

In this paper we consider theoretically the Kondo phenomenon in
electronic transport through two QDs coupled, in general, {\it
via} both Coulomb interaction and hopping term. To evaluate the
level renormalization and Kondo temperature of the system we use
the scaling approach. The Kondo temperature is also evaluated from
the slave boson technique. Additionally, the latter technique is
used to calculate the linear conductance. Then, the nonequllibrium
Green function formalism is used to calculate the local density of
states (LDOS) for both dots and transport characteristics
(differential conductance) in the nonlinear response regimes. To
calculate the relevant Green's functions from the corresponding
equations of motion we apply the decoupling scheme introduced in
Ref. [\onlinecite{meir2}].

The orbital Kondo effect in double quantum dot (DQD) systems was
analyzed e.g. in Ref. [\onlinecite{sztenkiel}]. However, our
results are different, and the key difference consists in a
different symmetry of the couplings to external leads. Moreover,
we use various techniques including scaling, slave boson, and
nonequilibrium Green function formalisms. Apart from this, we
apply a different method to evaluate the lesser Green function.
The paper is organized as follows. In section 2 we describe the
model of a double quantum dot system. Renormalization of the dots'
levels and the Kondo temperature are discussed in section 3 in
terms of the scaling approach. The slave boson technique is
briefly described in section 4 and is used there to estimate the
Kondo temperature and calculate the linear conductance. Basic
formula for nonequilibrium Green functions and the corresponding
numerical results on the conductance and LDOS are presented and
discussed in section 5. Summary and final conclusions are given in
section 6.

\section{Model}{\label{Sec:2}}

\begin{figure}
\begin{center}
\includegraphics[width=0.4\textwidth,angle=0]{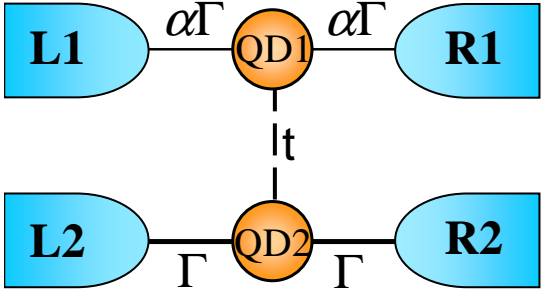}
\caption{\label{Fig:1}(color online) Schematic picture of the
double quantum dot system coupled to external leads. }
\end{center}
\end{figure}

We consider two coupled single-level quantum dots connected to
nonmagnetic electron reservoirs as shown schematically in
Fig.\ref{Fig:1}. Each dot is attached to separate source and drain
leads, so the tunneling paths {\it via} the two orbitals can be
analyzed separately as in recent experiments \cite{hubel,wilhelm}.
We consider the case when each dot is coupled symmetrically to the
leads, while the corresponding coupling strengths for both dots
may be different. Moreover, our considerations are limited to the
case of spinless electrons, which can be realized experimentally
for instance by applying a sufficiently strong external magnetic
field lifting the spin degeneracy.

The system under consideration can be described by the extended
Anderson Hamiltonian of the general form
\begin{equation}\label{1}
\hat{H}=\hat{H}_{\rm leads} +\hat{H}_{\rm DQD}+\hat{H}_{\rm
tunnel}.
\end{equation}
The first term, $\hat{H}_{\rm leads}$, describes here the four
leads in the non-interacting quasi-particle approximation,
$\hat{H}_{\rm
leads}=\hat{H}_{L1}+\hat{H}_{L2}+\hat{H}_{R1}+\hat{H}_{R2}$, with
$\hat{H}_{\beta i }$ being the Hamiltonian of the left ($\beta
=L$) and right ($\beta =R$) lead attached to the $i$th dot
($i=1,2$), $\hat{H}_{\beta i }=\sum_{\mathbf{k}}
\epsilon_{\mathbf{k}\beta i}c^{\dagger}_{\mathbf{k}\beta i}
c_{\mathbf{k}\beta i}$ (for $\beta ={\rm L,R}$ and $i=1,2$). Here,
$c^{\dagger}_{\mathbf{k}\beta i}$ ($c_{\mathbf{k}\beta i}$) is the
creation (annihilation) operator of an electron with the wave
vector $\mathbf{k}$ in the lead $\beta i$, whereas
$\epsilon_{\mathbf{k}\beta i}$ denotes the corresponding
single-particle energy.

The second term of the Hamiltonian (1) describes the double
quantum dot system,
\begin{eqnarray}
   \hat{H}_{DQD}=\sum_{i}\limits\epsilon_{i}d^\dag_{i}d_{i}+
   t(d^\dag_{1}d_{2}+h.c.)+
   Un_{1}n_{2},
\end{eqnarray}
where $n_{i}=d^\dag_{i}d_{i}$ is the particle number operator
($i=1,2$), $\epsilon_{i}$ is the discrete energy level of the
$i$-th dot, $t$ denotes the inter-dot hopping parameter (assumed
real), and $U$ is the inter-dot Coulomb integral.

The last term, $H_{\rm T}$, of Hamiltonian (1) describes electron
tunneling between the leads and dots, and takes the form
\begin{equation}
   \hat{H}_{\rm T}=\sum_{\mathbf{k}}\sum_{\beta i}
   (V_{i\mathbf{k}}^\beta c^\dag_{\mathbf{k}\beta i}d_{i}+\rm
   h.c.),
   \end{equation}
where $V_{i\mathbf{k}}^\beta$ are the relevant tunneling matrix
elements. Coupling of the dots to external leads can be
parameterized in terms of
$\Gamma^\beta_{i}(\epsilon)=2\pi\sum_\mathbf{k}
V_{i\mathbf{k}}^\beta
V^{\beta\ast}_{i\mathbf{k}}\delta(\epsilon-\epsilon_{\mathbf{k}\beta
i})$. We assume that $\Gamma^\beta_{i}$ is constant within the
electron band, $\Gamma^\beta_{i}(\epsilon)=\Gamma^\beta_{i}={\rm
const}$ for $\epsilon\in\langle-D,D\rangle$, and
$\Gamma^\beta_{i}(\epsilon)=0$ otherwise. Here, $2D$ denotes the
electron band width. We assume the dots are symmetrically coupled
to the leads, $\Gamma^L_{1}=\Gamma^R_{1}=\alpha\Gamma$, and
$\Gamma^L_{2}=\Gamma^R_{2}=\Gamma$. The parameter $\alpha$ takes
into account difference in the coupling of the two dots to
external leads. Note, that for these parameters, each dot
separately is coupled symmetrically to the two leads.

\section{Level renormalization and Kondo temperature }{\label{Sec:3}}

Coupling of the dots to external leads gives rise to
renormalization of the energy levels of both dots. In this section
we use the scaling approach to derive some general formula for the
renormalized levels in unbiased system. From the scaling equations
we also estimate the relevant Kondo temperature. The derived
results will be used subsequently for interpretation of the
numerical results on electronic transport and LDOS.

\subsection{Renormalization of the QDs' levels}

Now, we apply the scaling technique to derive renormalized dots'
energy levels and begin with the limit of $t=0$. In the scaling
approach, the high-energy excited states (in the energy region of
width $\delta D$ at the band edges) are removed, but their impact
on the system is taken into account {\it via} renormalized
parameters of the Hamiltonian. Here, we consider only second order
processes, where the leads' electrons are scattered to the band
edges and back. To perform scaling we assume $\epsilon_i+U\gg D\gg
|\epsilon_i|$. After integrating out the band edge states we
arrive at the following renormalized parameters;
\begin{equation}\label{SClev}
\tilde{\epsilon}_{i}=\epsilon_i-E_0+\frac{\Gamma_{\bar{i}}}{2\pi}\frac{\delta
D}{D},
\end{equation}
where $E_0$ is the energy of empty DQD system (initially $E_0=0$)
and the index $\bar{i}=1$ for $i=2$ and $\bar{i}=2$ for $i=1$.
Here, $\Gamma_{i}$ is defined as
$\Gamma_{i}=\Gamma^L_{i}+\Gamma^R_{i}$. This procedure leads to
the following scaling equation:
\begin{equation}\label{Eq:SC}
  \frac{d\tilde{\epsilon}_{i}}{d\ln D}=-\frac{\Gamma_{\bar{i}}}{2\pi},
\end{equation}
and to the level separation,
\begin{equation}\label{}
\Delta\tilde{\epsilon}=\tilde{\epsilon}_1-\tilde{\epsilon}_2=
\epsilon_1-\epsilon_2+\frac{(\Gamma_2-\Gamma_1)}{2\pi}{\rm
ln}\left(\frac{D}{\widetilde{D}}\right),
\end{equation}
where $\widetilde{D}$ is the band width at the end of scaling
procedure. When $\epsilon_1=\epsilon_2=\epsilon_0$, the above
equation shows that the initial degeneracy is generally lifted.

Scaling in the presence of direct tunneling between the dots,
$t\neq 0$, is more complex in a general case. However, we restrict
our considerations to some limiting cases, i.e. when the hoping
term is weak, $|t|/\Gamma \ll 1$, and when  $|t|/\Gamma \gg 1$. If
the bare dots' levels are degenerate, the direct hopping term
generally lifts the degeneracy. The two eigenstates of the coupled
quantum dots isolated from the leads correspond to the antibonding
and bonding states, with the eigenenergies
$\epsilon_{\pm}=(\epsilon_1+\epsilon_2)/2\pm\sqrt{\Delta\epsilon^2
+t^2}$, where $\Delta\epsilon=(\epsilon_1-\epsilon_2)/2$. When
the hoping term is small, one can first perform scaling of the
bare dots' levels and then incorporate nonzero $t$ by substituting
$\epsilon_{1(2)}$ by $\tilde{\epsilon}_{1(2)}$ in the above
expression for $\epsilon_{\pm}$. The situation changes when
tunneling coupling between the dots is larger than the dot-lead
coupling. To find the relevant energy levels involved in the Kondo
effect, one has to diagonalize the dot's Hamiltonian first
(transformation to the bonding and antibonding states), and then
perform scaling for the energy levels $\epsilon_{+}$ and
$\epsilon_{-}$. The corresponding scaling equation has the form
\begin{equation}\label{SCpm}
  \frac{d\tilde{\epsilon}_{\pm}}{d\ln
  D}=-\frac{\Gamma_{\mp}}{2\pi},
\end{equation}
which is similar to Eq.(\ref{Eq:SC}). However, the effective coupling
of the new states to the leads acquires now the form
$\Gamma_{\pm}=\Gamma_1+\Gamma_2$. Thus, the level separation in
the limit of strong hoping term is independent of the couplings
$\Gamma_{\pm}$ as they are both the same, $\Gamma_{+}=\Gamma_{-}$.

\subsection{Kondo temperature}

Now we evaluate the Kondo temperature  for $t=0$ and
$\epsilon_1=\epsilon_2=\epsilon_0$ using the 'poor man' scaling
approach\cite{hewson}. To pursue this method one has to derive
first the Kondo Hamiltonian by performing the Schrieffer-Wolf
transformation. Then, the band width is reduced by eliminating
states with energy $D-\delta D\leq|\epsilon_{\mathbf{k}\beta
i}|\leq D$ and introducing a new effective Kondo Hamiltonian,
which has the same form as the initial one, but with renormalized
parameters $\tilde{J}_+,\tilde{J}_-,\tilde{J}_{z1}$ and
$\tilde{J}_{z2}$. All information on the high energy excitations
is incorporated into these renormalized parameters.

To apply the renormalization group procedure we first reformulate
the definitions of the coupling parameters $\Gamma_{i}$ in the
following way: $\Gamma_{1}=2\alpha\Gamma
\equiv\tilde{\Gamma}(1-\tilde{p})$ and $\Gamma_{2}=2\Gamma
\equiv\tilde{\Gamma}(1+\tilde{p})$, so that
$\tilde{\Gamma}=(\alpha+1)\Gamma $ and
$\tilde{p}=(1-\alpha)/(1+\alpha)$. After performing scaling
procedure one arrives at the following scaling equations;
\begin{equation}\label{RG1}
    \frac{d(\rho_{}\tilde{J}_{\pm})}{d\ln D}=-\rho_{}\tilde{J}_{\pm}(\rho\tilde{J}_{z1}+\rho\tilde{J}_{z2}),
\end{equation}
\begin{equation}\label{RG2}
\frac{d(\rho\tilde{J}_{zi})}{d\ln D}=-2(\rho\tilde{J}_{\pm})^2,
\end{equation}
where $\rho=\rho_1=\rho_2$, with $\rho_i=\sum_{\beta}\rho_{\beta
i}$, and $\rho_{\beta i}$ being the density of states in the lead
$\beta i$ ($\beta=L, R$ and $i=1,2$).  To solve these equations we
first find the scaling trajectories
$(\rho_{}\tilde{J}_{\pm})^2-(\rho_{}\tilde{J}_{z1})(\rho_{}\tilde{J}_{z2})=0$
and $\rho_{}\tilde{J}_{z1}-\rho_{}\tilde{J}_{z2}=const\equiv\rho
J_{z1}^0-\rho J_{z2}^0=\tilde{p}\rho (J_{z1}^0+J_{z2}^0)$. This
allows us to write only one scaling equation instead of the two
coupled equations (\ref{RG1}) and (\ref{RG2}),
\begin{equation}\label{RG}
    \frac{d(\rho\tilde{J}_{zi})}{d\ln D}=-2(\rho_{}\tilde{J}_{zi})[\rho_{}\tilde{J}_{zi}\mp
    \rho\tilde{p}(J_{z1}^0+J_{z2}^0)],
\end{equation}
for $i=1,2$. One actually continues the scaling process until
$D\approx k_BT_K$. Solving Eq.~(\ref{RG}) one finds the Kondo
temperature as the relevant scaling invariant,
\begin{equation}\label{T_K}
    T_K=\widetilde{D}\exp\left\{-\frac{1}{\rho (J_{z1}^0+J_{z2}^0)}\frac{{\rm
    arctanh}(\tilde{p})}{\tilde{p}}\right\},
\end{equation}
with
$\rho(J_{z1}^0+J_{z2}^0)=\frac{2\widetilde{\Gamma}}{\pi}\frac{U}{|\epsilon_0|(\epsilon_0+U)}$.
The above formula resembles the corresponding one for the Kondo
temperature in a single QD coupled to ferromagnetic
leads\cite{martinek}, where instead of $\tilde{p}$ we have spin
polarization $p$ of the leads.

\begin{figure}
\begin{center}
\includegraphics[width=0.46\textwidth]{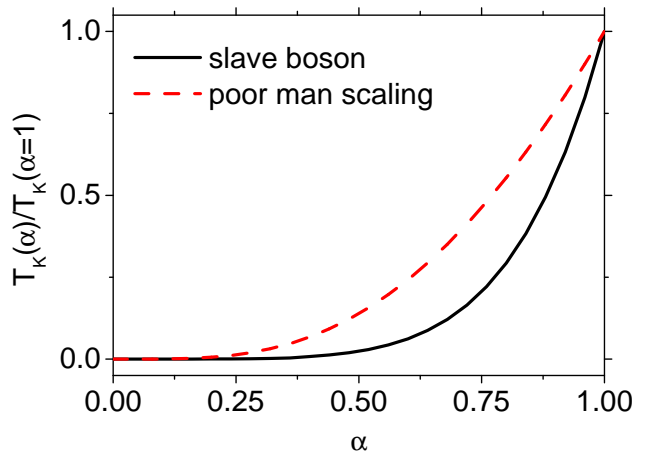}
\caption{\label{Fig:2}Normalized Kondo temperature as a function
of the parameter $\alpha$, obtained from the scaling method for
$U=50\Gamma$ and from the slave boson technique for $U=\infty$.
The other parameters are $\epsilon_0=-3.5\Gamma$ and $t=0$.}
\end{center}
\end{figure}

Variation of the Kondo temperature with the parameter $\alpha$ is
shown in Fig.\ref{Fig:2} (dashed line). $T_K$ reaches maximum
value for $\alpha=1$ ($\tilde{p}=0$) and vanishes for
$\alpha\rightarrow 0$ ($\tilde{p}=1$). This behavior is similar to
that for spin Kondo phenomenon in a QD coupled to ferromagnetic
leads. We also note that our results are in agreement with those
obtained in Ref.[\onlinecite{wohlmann}], where the authors mapped
spinless DQD system onto a spinful generalized Anderson model.

\section{Slave boson approach}{\label{Sec:4}}

To estimate the Kondo temperature and calculate conductance in the
linear response regime, we apply now the slave boson technique for
$U\rightarrow\infty$\cite{coleman}. This method relies on
introducing auxiliary operators for the dots, and replacing the
electron creation and annihilation operators by $f^{\dag}_{i}b$
and $b^{\dag}f_{i}$, respectively. Here, $b^{\dag}$ creates an
empty state, whereas $f^{\dag}_{i}$ creates a singly occupied
state with an electron in the $i$-th dot. To eliminate
non-physical states, the following constraint has to be imposed on
the new quasi-particles,
\begin{equation}\label{Eq:constraint}
    Q=\sum_{i}\limits
    f^{\dag}_{i}f_{i}+b^{\dag}b=1.
\end{equation}
The above constraint prevents double occupancy of the system (the
DQD system is either empty or singly occupied).

In the mean field approximation (MFA), the boson field $b$ is
replaced by an independent of time real number,
$b(t)\rightarrow\langle b(t)\rangle\equiv{\bar b}$. This
approximation, however, restricts considerations to the low bias
regime ($eV\ll |\epsilon_i|$). Introducing now the following
renormalized parameters: $\bar{t}=t\bar{b}^2$,
$\overline{V}_{i\mathbf {k}}^\beta =V_{i\mathbf
{k}}^\beta\bar{b}$, and $\bar{\epsilon}_{i}=\epsilon_{i}+\lambda$,
where $\lambda$ is the corresponding Lagrange multiplier, one can
write the effective MF Hamiltonian as
\begin{align}\label{Eq:HMFA}
\hat{H}^{MF}=&\sum_{\mathbf k} \sum_{\beta i}\varepsilon_{{\mathbf
k}\beta i}c^{\dagger}_{{\mathbf k}\beta i} c_{{\mathbf k}\beta i}+
\sum_{i}\limits\bar{\epsilon}_{i}f^\dag_{i}f_{i}+(\bar{t}f^\dag_{1}f_{2}+{\rm h.c.})
  \notag \\
   +&\sum_{{\mathbf k}}\sum_{\beta i}\limits
   (\overline{V}_{i\mathbf{k}}^\beta
   c^\dag_{{\mathbf k}\beta i}f_{i}+{\rm h.c.})
  +  \lambda\left(\bar{b}^2-1\right).
\end{align}
The unknown parameters, $\bar{b}$ and $\lambda$, have to be found
self-consistently from the following equations;
\begin{equation}\label{}
    \bar{b}^2-i\sum_{i}\limits
    \int\frac{d\varepsilon}{2\pi}\langle\langle
    f_{i}|f_{i}^{\dag}\rangle\rangle^<_\varepsilon =1,
\end{equation}
\begin{eqnarray}\label{}
-i\sum_{i}\limits\int\frac{d\varepsilon}{2\pi}(\varepsilon-\bar{\epsilon}_{i})
\langle\langle f_{i}|f^{\dag}_{i}\rangle\rangle^<_\varepsilon +
\lambda\bar{b}^2=0,
\end{eqnarray}
where $\langle\langle
    f_{i}|f^{\dag}_{i}\rangle\rangle^<_\varepsilon$
    is the Fourier transform of the
    lesser Green function defined as $G^<_{ii}(t,t^\prime)\equiv \langle\langle
    f_{i}(t)|f^{\dag}_{i}(t')\rangle\rangle^<=i
    \langle
    f^{\dag}_{i}(t')f_{i}(t)\rangle$.
These equations follow from the constraint imposed on the slave
boson field, Eq.~(\ref{Eq:constraint}), and from the equation of
motion for the slave boson operator. The lesser Green functions
$\langle\langle f_{i}|f^{\dag}_{i}\rangle\rangle^<_\varepsilon$ as
well as the retarded Green functions $\langle\langle
f_{i}|f^{\dag}_{i}\rangle\rangle^r$ (the latter ones are required
in the further calculations, too) have been determined from the
corresponding equations of motion.

The Kondo temperature can be introduced as\cite{lim}
\begin{equation}\label{Kondo}
    T_K\equiv\sqrt{\bar{\epsilon}_0^2+\overline{\Gamma}^2},
\end{equation}
with $\overline{\Gamma}=\bar{b}^2(\Gamma_1+\Gamma_2)$ and
$\overline{\epsilon}_1=\overline{\epsilon}_2=\overline{\epsilon}_0$.
Variation of the Kondo temperature (evaluated from the above
equation) with the parameter $\alpha$ is shown in Fig.\ref{Fig:2} (solid
line).

\begin{figure}
\begin{center}
\includegraphics[width=0.46\textwidth]{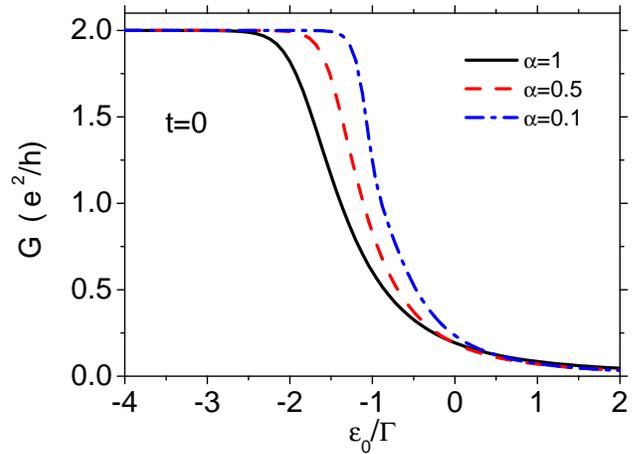}
\caption{\label{Fig:3}Linear conductance vs. dots' level position,
$\epsilon_1=\epsilon_2=\epsilon_0$, for indicated values of
$\alpha$ and $t=0$, obtained from the slave boson method for
$U\to\infty$. The level splitting due to renormalization is tuned
out by external gates. }
\end{center}
\end{figure}

To study charge transport we assume the same electrochemical
potentials for the left leads and also equal electrochemical
potentials of the right leads.  The linear conductance is then
calculated from the Landauer formula, in which the transmission
matrix is taken at the Fermi level. More specifically linear conductance is given
by the formula
\begin{equation}\label{Eq:lc}
G_{V\rightarrow 0}=\lim_{V\rightarrow 0}\frac{d J}{dV},
\end{equation}
where $J$ is current calculated at temperature $T=0K$\cite{trochaJNN}.
The slave boson technique in
the form presented above, however, does not take into account the
level renormalization described in the preceding sections.
Therefore, to include this renormalization  we replace the bare
dot levels by the renormalized ones (keeping the notation used for
the bare dot levels). Alternatively one may say that the
renormalization is tuned out by external gate voltages. In
Fig.\ref{Fig:3} the linear conductance is shown as a function of
the dots' energy level, $\epsilon_1=\epsilon_2=\epsilon_0$, and
for indicated values of the parameter $\alpha$. The linear
conductance reaches the unitary limit for $\epsilon_0\ll -\Gamma$.
This limit is achieved owing to the tuning out the level splitting
due to renormalization. From this figure also follows, that the
Kondo temperature decreases with decreasing $\alpha$, in agreement
with the above discussion and Fig.2.

\section{Non-equilibrium Green function approach}{\label{Sec:5}}

Electric current flowing through a biased system is determined by
nonequillibrium retarded, advanced, and lesser Green functions of
the dots, and can  be calculated from the  formula derived by Meir
et al \cite{meir92}. In turn, to calculate the retarded (advanced)
Green functions $G^{r(a)}_{ij}(\epsilon)$, we have applied the
equation of motion method (EOM). Within this method one writes first the
equation of motion for the causal Green function
$G_{ij}(\epsilon)$, which generates new Green functions. Then, one
writes the equations of motion for these new Green functions,
which in turn contain new higher-order Green functions. The latter
ones have to be calculated approximately. To close the set of
equations for the Green functions we have applied the decoupling
scheme introduced in Ref.[\onlinecite{meir2}]. Although such an
approximation does not describe properly the zero temperature
limit, it is sufficient to describe the Kondo phenomenon close to
the Kondo temperature. Detailed expressions for these Green functions
are shown in the Appendix.

The retarded/advanced Green functions contain occupation numbers,
$n_{i}$, and the interdot correlators, $n_{i\bar{i}}=\langle
d^\dag_{i}d_{\bar i}\rangle$, which can be calculated from the
identities
\begin{align}\label{Eq:n1}
    n_{i}&=-i\int\frac{d\epsilon}{2\pi}G^<_{ii}(\epsilon),
\end{align}
\begin{align}\label{Eq:n2}
    n_{i\bar{i}}&=-i\int\frac{d\epsilon}{2\pi}G^<_{\bar{i}i}(\epsilon).
\end{align}
Thus, we still need the lesser Green functions $G_{ij}^<(\epsilon
)$.  However, one can note that instead of $G_{ij}^<(\epsilon)$,
only $\int d\epsilon \, G_{ij}^<(\epsilon)$ is needed. This
quantity can be found exactly (in contrast to the approach based
on the Ng's approximation\cite{Ng}) and to do this we apply the
Heisenberg equation of motion for the  operators
$d^\dag_{j}(t)d_{i}(t)$. Then, one takes average from the obtained
equation and makes use of the fact that $\langle (d/dt)
d^\dag_{j}(t)d_{i}(t)\rangle =0$ in a steady state. As a result
one obtains the following equations;
\begin{align}\label{Eq:SC1}
   t(n_{i\bar{i}}- n_{\bar{i}i}) -i\Gamma_{i}n_i
    \nonumber
   \\
   =\sum_\beta\limits\int\frac{d\epsilon}{2\pi}
    f_{\beta}(\epsilon)\Gamma_{i}^\beta(G_{ii}^{r}-G_{ii}^{a})&,
\end{align}
\begin{align}\label{Eq:SC2}
   t(n_{i}- n_{\bar{i}})+
   (\epsilon_{\bar i}-\epsilon_i)n_{i{\bar i}}-
   \frac{i}{2}(\Gamma_{i}n_{i{\bar i}}+
   \Gamma_{{\bar i}}n_{i{\bar i}})
    \nonumber
    \\
   =\sum_\beta\limits\int\frac{d\epsilon}{2\pi}
    f_{\beta}(\epsilon)[\Gamma_{i}^\beta G^{r}_{\bar{i}i}-
    \Gamma_{{\bar i}}^{\beta}G_{{\bar i}i}^{a}]&.
\end{align}
These equations, together with the appropriate equations for the
retarded/advanced Green functions, have to be solved numerically
in a self-consistent way.

\begin{figure}
\begin{center}
\includegraphics[width=0.46\textwidth]{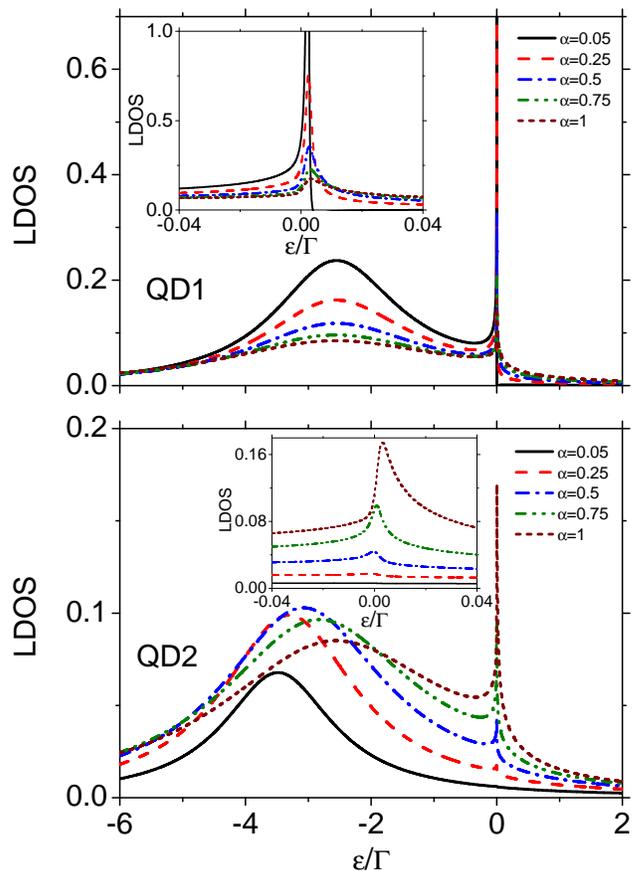}
\caption{\label{Fig:4}Local density of states for the dots QD1 and
QD2, calculated as a function of energy for indicated values of
the parameter $\alpha$. The other parameters are
$\epsilon_{0}=-3.5\Gamma$, $U=50\Gamma$ and $t=0$.}
\end{center}
\end{figure}

The basic transport characteristics of the system, like
conductance and differential conductance can be calculated
numerically using the formulas derived above. The local density of
states (LDOS) for the $i$-th dot can be calculated as
\begin{equation}\label{}
   D_i=-\frac{1}{\pi}\Im\left[G_{ii}^r(\varepsilon)\right],
\end{equation}
where $\Im[A]$ denotes the imaginary part of $A$.

The approximation scheme used to calculate the nonequillibrium
Green functions does not take into account the level
renormalization described in the preceding sections. Therefore, to
take this renormalization into account we replace the bare dot
levels by the renormalized ones (keeping the notation used for the
bare dot levels), similarly as in the case of slave boson
technique. However, one should bear in mind that the presented EOM approach just renormalizes the bare dot's energy levels due to real part of the corresponding self-energies. This renormalization can be seen looking at the position of the broad maximum in LDOS (see Fig.\ref{Fig:4}). However, the used decoupling scheme does not take it properly and thus does not lead to the expected splitting of the zero bias anomaly.
In the following numerical calculations we assume equal
dot energy levels, $\epsilon_{i}=\epsilon_0$ (for $i =1,2$)
($\epsilon_0$ is measured from the Fermi level of the leads in
equilibrium, $\mu_{Li}=\mu_{Ri}=0$). Apart from this,  we assume
$\epsilon_{0}=-3.5\Gamma$, the bandwidth $2D=500\Gamma$, and
$U=50\Gamma$.

\subsection{Numerical results for  $t=0$}

Let us start with the case when the dots are  capacitively coupled
only, $t=0$. The LDOS  for both dots is plotted in
Fig.\ref{Fig:4}. The spectrum of each dot reveals two resonances
corresponding to the  dot level and its Coulomb counterpart (the
latter not shown). Apart from this, a narrow peak emerges in the
spectrum of each dot at the Fermi level of the leads. The
intensity and width of this peak strongly depends on temperature,
revealing all characteristic features typical of the Kondo
resonance.

The resonance in LDOS originates from the many body processes
which occur in the low temperature regime. Since  the conditions
$\epsilon_0<\mu_{\beta i}$ and $\mu_{\beta i} <\epsilon_0+U$ are
obeyed for the parameters assumed (Coulomb blockade regime), only
a single electron can occupy the DQD system and sequential
tunnelling processes are blocked. However, higher-order tunnelling
events are still allowed. Let us assume that an electron initially
occupies the dot QD1, and the system is in the Coulomb blockade
regime. Due to the uncertainty principle, the electron from the
dot QD1 can tunnel onto the Fermi level of one of the leads
attached to QD1, while an electron from the Fermi level of one of
the leads attached to QD2 can tunnel to the dot QD2 in the time
$\hbar /|\epsilon_0|$. Interference of many such events gives rise
to the narrow peaks in LDOS at the Fermi level.

For the fully symmetric model ($\alpha=1$), LDOS for the dot QD1
is the same as that for QD2. The situation is different for
$\alpha\neq 1$. As $\alpha$ decreases, the intensity of the Kondo
peak in LDOS of the dot QD2 also decreases and disappears when
$\alpha$  tends to zero. The opposite situation occurs in the LDOS
of the dot QD1, where the Kondo peak becomes more and more
pronounced with decreasing $\alpha$. This behavior is due to the
fact that the intensity of the Kondo peak in LDOS of the dot QD1
is mainly determined by the coupling strength between QD2 and the
leads, while the Kondo peak for the dot QD2 is predominantly
determined by coupling of the dot QD1 to the leads.
Accordingly,
the Kondo peak in LDOS of the dot QD1 (QD2) increases (decreases)
with decreasing $\alpha$, while the Kondo peaks of both dots are
equal for $\alpha=1$.
This can be understood taking into account the maximal value of the conductance per each channel (each dot), which is equal to $G^{max}_{i}=e^2/h$. Estimating $G^{max}_{i}\sim\Gamma_{ii}(E_F)\rho_i(E_F)$ at the Fermi level, one obtains $\rho_1(E_F)/\rho_2(E_F)=1/\alpha$, which explains the above behavior of the LDOS for QD1 and QD2.
This behavior of the Kondo peaks in LDOS of
both dots is similar to that in the case of spin Kondo effect,
where each spin channel is coupled differently to the leads (when
the leads are ferromagnetic).

For $\alpha \ll 1$, the Kondo peak for the dot QD1 becomes
strongly asymmetric and the LDOS is totally suppressed for
energies above the Fermi level, where the spectral function is
equal to zero. This situation is similar to that reported in
Ref.[\onlinecite{sztenkiel2}]. Apart from this, position of the
Kondo peak for QD2 slightly moves away from the Fermi level with
increasing $\alpha$ (towards positive energies), and becomes
asymmetric for all values of $\alpha$.

It is also worth to note that position of the broad maximum
(associated with the dot's level) in LDOS of QD1 (the dot whose
coupling to the leads changes with $\alpha$) is almost unchanged
with tunning $\alpha$, whereas position of the broad maximum in
the LDOS of the dot QD2 (coupled to the leads with constant
strength) varies with the parameter $\alpha$. This becomes clear when
considering the formulas for renormalized dots' energy levels, Eq.(\ref{Eq:SC}).
 Apart from this, the
intensity of the broad peak for the dot QD1 decreases
monotonically with increasing $\alpha$, whereas the intensity of
the broad peak in the LDOS of the dot QD2 depends on $\alpha$ in a
more complex way. When $\alpha\rightarrow 0$, the broad peak in
the LDOS of the dot QD1 is then  most pronounced, whereas its
Coulomb counterpart (not shown) is totally localized at
$\epsilon_0+U$.

\begin{figure}
\begin{center}
\includegraphics[width=0.48\textwidth]{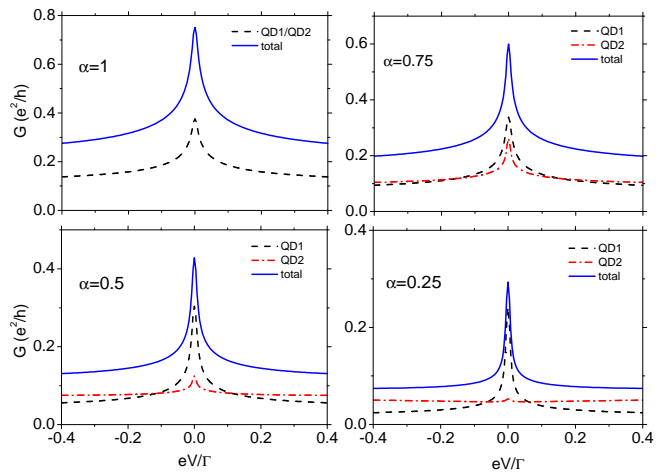}
\caption{\label{Fig:5}Differential conductance for the dots QD1
and QD2, and the total differential conductance calculated for
indicated values of $\alpha$. The other parameters as in
Fig.\ref{Fig:4}.}
\end{center}
\end{figure}

As mentioned before, the resonances in LDOS lead to zero bias
anomaly in the differential conductance of the  DQD system.
Here, nonlinear conductance is defined in the following way:
\begin{equation}\label{Eq:nlc}
G=\frac{d J}{dV},
\end{equation}
where $J$ is current given by Meir and Wingreen formula\cite{meir92}.
This quantity is very important from practical point of view as it is usually measured
in QDs' experiments (to obtain basic transport properties of these systems)\cite{cronewett}.
In Fig.(\ref{Fig:5}) we show the differential conductance of both
dots as a function of the bias voltage. For a fully symmetric
system, the differential conductance of both dots is the same, but
the situation changes when $\alpha$ becomes smaller than 1,
$\alpha < 1$. Interestingly, the conductance of the dot weakly
coupled to the leads (in our case the dot QD1) is larger than the
conductance of the dot strongly coupled to the leads. For a
sufficiently small value of $\alpha$, the differential conductance
of the dot QD2 appears as a broad background, whereas the
differential conductance of the dot weakly coupled to the leads is
then very narrow. This behavior follows the features of LDOS of
the dots QD1 and QD2 discussed above (Fig.\ref{Fig:4}).
\begin{figure}
\begin{center}
\includegraphics[width=0.46\textwidth]{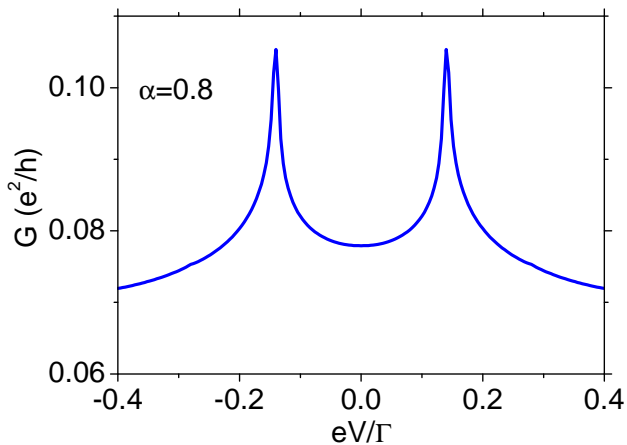}
\caption{\label{Fig:5b} Differential conductance calculated for
indicated value of $\alpha$ in the case when the level slitting due to
renormalization is not compensated by external gate voltages. The positions
of the dots' levels have been estimated self-consistently using Eq.(\ref{Eq:SC}).
 The other parameters as in
Fig.\ref{Fig:4}.}
\end{center}
\end{figure}
Apart from this, we note that the total differential conductance
diminishes as $\alpha$ decreases, and its line width also shrinks.
This behavior indicates on the suppression of the effective Kondo
temperature as $\alpha$ decreases. Such a behavior stems from the
fact that the rate of tunneling events leading to the Kondo
resonance decreases since the dot QD1 becomes detached from the
leads as $\alpha$ decreases. This is also in agreement with our
predictions on the $\alpha$ dependence of the Kondo temperature,
derived in Sections \ref{Sec:3} and \ref{Sec:4} (see also
Fig.\ref{Fig:2}). Finally, when one of the dot is totally
disconnected from the leads ($\alpha=0$), the Kondo temperature
vanishes, and no Kondo effect appears.

When $\epsilon_1\ne\epsilon_2$ (eg. when the level slitting due to
renormalization is not compensated by external gate voltages), the
Kondo peak in differential conductance becomes split and the two
components are shifted from the Fermi level and have rather low
intensity, as shown in Fig.\ref{Fig:5b}. This suppression of the
Kondo anomaly resembles similar behavior in the case of the spin
Kondo effect.

The presence of Kondo peaks also depends on the coupling strength
of the dots to the leads. Above we assumed relatively strong
coupling for both dots, with some asymmetry of this coupling
described by the parameter $\alpha$. We have also examined the
case when both dots are weakly coupled to the leads. There is no
Kondo effect in such a case, which is consistent with the recent
experimental observations \cite{hubel}.

\begin{figure}
\begin{center}
\includegraphics[width=0.46\textwidth]{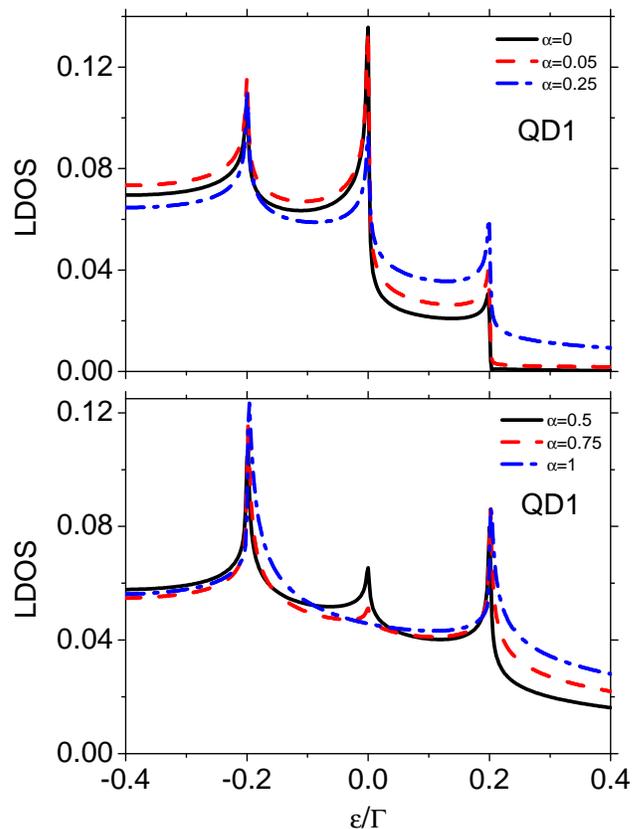}
\caption{\label{Fig:6} Local density of states for the dot QD1,
calculated for indicated values of $\alpha$ and for
$t=-0.1\Gamma$. The other parameters as in Fig.\ref{Fig:4}.}
\end{center}
\end{figure}

\begin{figure}
\begin{center}
\includegraphics[width=0.46\textwidth]{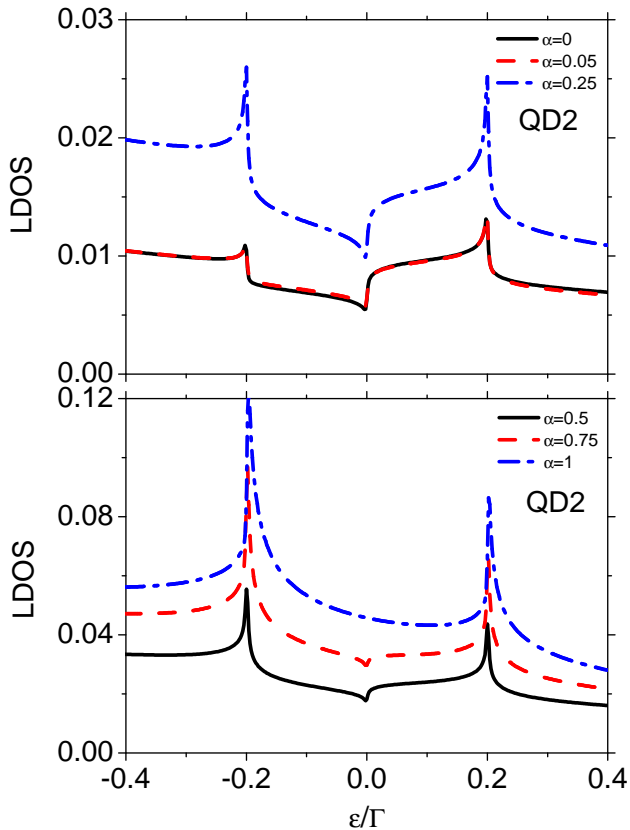}
\caption{\label{Fig:7} Local density of states for the dot QD2,
calculated for indicated values of $\alpha$ and for
$t=-0.1\Gamma$. The other parameters as in Fig.\ref{Fig:4}.}
\end{center}
\end{figure}

\subsection{Numerical results for the case $t\neq 0$}

Now, we consider the situation when direct hopping between the
dots is allowed. Figures \ref{Fig:6} and \ref{Fig:7} show LDOS for
the dots QD1 and QD2. When both dots are equally coupled to the
leads ($\alpha=1$), a double peak structure emerges in the LDOS of
the dots QD1 and QD2, and the LDOS is the same for both dots. The
two peaks are centered at  $\varepsilon=\pm 2t$. This comes from
the fact that when $t\neq 0$, the dots' states hybridize into two
molecular-like states with eigenenergies
$\epsilon_{\pm}=\epsilon_0\pm t$. These new states are then
involved in the Kondo phenomenon. If initially an electron
occupies the level $\epsilon_{-}$, then it can tunnel into the
Fermi level of a given lead and simultaneously another electron
 having energy $+2t$  tunnels onto
the level $\epsilon_{+}$. Coherent superposition of many such
events results in the Kondo peak at the energy $\varepsilon=2t$.
In the same way one may explain the presence of the Kondo peak at
the energy $\varepsilon=-2t$.

However, the situation changes  for $\alpha\neq 1$. Apart from the
two peaks located at $\pm 2t$, one finds an additional peak in the
LDOS of the dot QD1, which is located at the Fermi level. However,
instead of the peak, a dip in the LDOS of the dot QD2 appears at
the Fermi level. Possible explanation of this behavior relies on
the transitions/tunneling events which do not induce electron
exchange between the molecular states $\epsilon_{+}$ and
$\epsilon_{-}$, but rather between original bare dot levels.
The appearance of the dip (at the Fermi level) can be explained using the arguments from Section \ref{Sec:5} A. One can notice that for $\alpha=1$ (symmetric couplings) there is no peak (or dip) in the LDOS at the Fermi level and LDOS for QD1 and QD2 are equal. For asymmetric couplings, i.e., $\alpha\neq 1$, the amplitude of the LDOS at the Fermi level for the dot weakly coupled to the leads exceeds that for the dot strongly coupled to the leads. As a result, the amplitude of the LDOS at the Fermi level for the dot strongly coupled to the leads decreases (as $\alpha$ decreases) and the dip structure occurs.

\begin{figure}
\begin{center}
\includegraphics[width=0.46\textwidth]{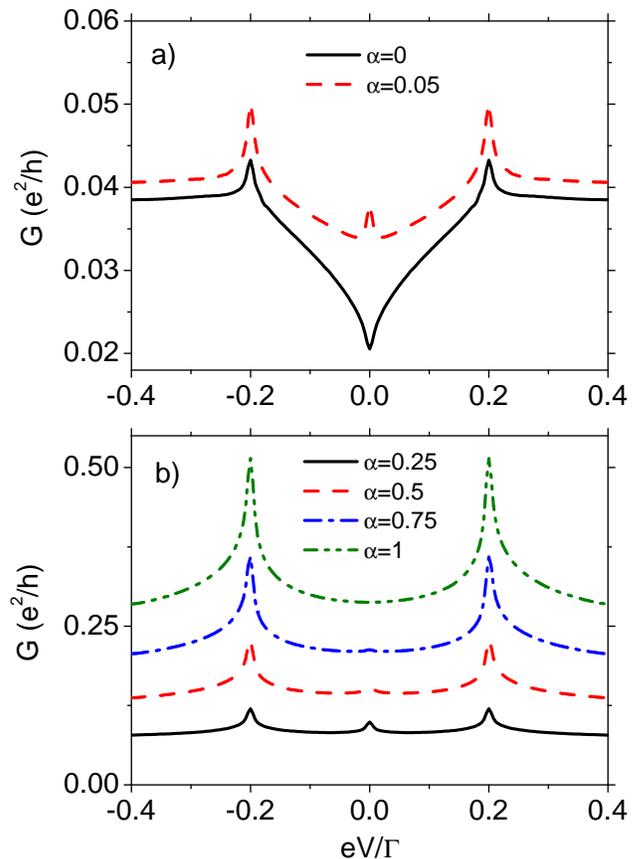}
\caption{\label{Fig:8} Differential conductance for the dots QD1
and QD2, and total differential conductance calculated for
indicated values of $\alpha$ and for $t=-0.1\Gamma$. The other
parameters as in Fig.\ref{Fig:4}.}
\end{center}
\end{figure}

Differential conductance for several values of the asymmetry
parameter $\alpha$ is shown in Fig.\ref{Fig:8}. When the dots are
connected in the T-shape geometry, $\alpha=0$, one finds two
maxima centered at $eV=\pm 2t$, and one dip at $eV=0$. Suppression
of the conductance at $eV=0$ is a result of destructive quantum
interference\cite{trocha}. When the coupling to the dot QD1 is
turned on, then the dip structure disappears in the differential
conductance. Instead of dip, one finds the third peak centered at
$eV=0$. For a fully symmetric system, $\alpha=1$, this peak
vanishes and only the satellite maxima are present. It is also
worth to note that the conductance increases with increasing
$\alpha$.

\section{Summary and conclusions}

We have considered the orbital Kondo effect in a spinless system
of two single-level quantum dots connected to electron reservoirs.
Various techniques have been used to describe basic features of
the Kondo physics. First, we used the scaling technique to
evaluate the level renormalization and the corresponding Kondo
temperature. Then, we used the slave boson technique to calculate
local density of states and linear conductance. To find nonlinear
conductance we used the nonequillibrium Green function method.

The numerical results show that transport characteristics reveal
typical Kondo phenomenon, similar to that observed in a single
quantum dot with spin degenerate discrete level, coupled to
external ferromagnetic leads. In the case considered, the
splitting due to level renormalization could be compensated by
external gate voltages, so one could reach the full Kondo anomaly.
Such a compensation, however, is not possible when the direct
tunneling between the dots is strong.

\begin{acknowledgements}
The author is extremely gratefull to Prof. J. Barna\'s for fruitful discussions and
would like to thank him for giving helpful advices during preparation of the manuscript.
This work, as part of the European Science Foundation EUROCORES
Programme SPINTRA, was supported by funds from the Ministry of
Science and Higher Education as a research project in years
2006-2009 and the EC Sixth Framework Programme, under Contract N.
ERAS-CT-2003-980409.
The author also acknowledges support by funds from Ministry of Science and Higher
Education as a research project N N202 169536 in years 2009-2011.
\end{acknowledgements}

\appendix*

\section{Green's functions}
Here we show explicit form of the derived dots' Green functions
$G_{ij}$ for $i, j=1, 2$;

\begin{widetext}

\begin{eqnarray}\label{}
G_{ii}=\frac{1}{M}\left\{\left[1+\frac{U}{W}\left(n_{\bar
i}A_{\bar{i}}-n_{i\bar{i}}\tilde{t}\right)\right]\Omega_{\bar{i}\bar{i}}
+
    \frac{U}{W}\left(n_{\bar i}\tilde{t}-n_{i\bar{i}}A_i\right)\Omega_{i\bar{i}}\right\},
\end{eqnarray}
\begin{eqnarray}\label{}
 G_{i\bar{i}}=\frac{1}{M}\left\{\frac{U}{W}\left(n_{i}\tilde{t}-n_{\bar{i}i}A_{\bar{i}}\right)\Omega_{\bar{i}\bar{i}}
 +
 \left[1+\frac{U}{W}\left(n_iA_i-n_{\bar{i}i}\tilde{t}\right)\right]\Omega_{i\bar{i}}\right\},
\end{eqnarray}

\end{widetext}

 where
  \[M=\Omega_{11}\Omega_{22}-\Omega_{12}\Omega_{21}\]
    \[\Omega_{ii}=\epsilon-\epsilon_i-\Sigma_{ii}^{(0)}+\frac{U}{W}\left(\alpha_i A_{\bar{i}}+\gamma_{\bar i}\tilde{t}\right)\]
    \[\Omega_{i\bar{i}}=t-\frac{U}{W}\left(\gamma_i A_{\bar{i}}+\alpha_{\bar i}\tilde{t}\right),\]
    \[W=A_1A_2-\tilde{t}^2\]
\[A_1=\epsilon-\epsilon_1-U-\Sigma_{11}^{(0)}-\Sigma_{11}^{c}-\Sigma_{22}^{e}-\Sigma_{22}^{d},\]
\[A_2=\epsilon-\epsilon_2-U-\Sigma_{22}^{(0)}-\Sigma_{22}^{c}-\Sigma_{11}^{f}-\Sigma_{11}^{d},\]
\[\tilde{t}=t+\Sigma_{22}^{a}+\Sigma_{11}^{b},\]
\[\alpha_1=\Sigma_{22}^{dI}+\Sigma_{11}^{cI}+\Sigma_{22}^{eI},\]
\[\alpha_2=\Sigma_{11}^{dI}+\Sigma_{11}^{fI}+\Sigma_{22}^{cI},\]
\[\gamma_1=\Sigma_{22}^{aI}+\Sigma_{11}^{bI},\]
\[\gamma_2=\Sigma_{22}^{aI}+\Sigma_{11}^{bI}.\]

The self-energies are defined in the following way
\[\Sigma_{ii}^{a(n)}=\sum_{\mathbf{k}\alpha}\limits
|V_{i\mathbf{k}}^{\alpha}|^2
\frac{t}{\Lambda}(\epsilon-\epsilon_{\mathbf{k}\alpha}+\Delta\epsilon)F_\alpha^{(n)}(\epsilon_{\mathbf{k}\alpha})
\]
\[\Sigma_{ii}^{b(n)}=\sum_{\mathbf{k}\alpha}\limits
|V_{i\mathbf{k}}^{\alpha}|^2
\frac{t}{\Lambda}(\epsilon-\epsilon_{\mathbf{k}\alpha}-\Delta\epsilon)F_\alpha^{(n)}(\epsilon_{\mathbf{k}\alpha})
\]
\[\Sigma_{ii}^{c(n)}=\sum_{\mathbf{k}\alpha}\limits
|V_{i\mathbf{k}}^{\alpha}|^2
\frac{2t^2}{\Lambda}F_\alpha^{(n)}(\epsilon_{\mathbf{k}\alpha})
\]
\[\Sigma_{ii}^{d(n)}=\sum_{\mathbf{k}\alpha}\limits
\frac{|V_{i\mathbf{k}}^{\alpha}|^2}{\epsilon+\epsilon_{\mathbf{k}\alpha}-\epsilon_1-\epsilon_2-U}
F_\alpha^{(n)}(\epsilon_{\mathbf{k}\alpha})
\]

\begin{widetext}

\[\Sigma_{ii}^{e(n)}=\sum_{\mathbf{k}\alpha}\limits
|V_{i\mathbf{k}}^{\alpha}|^2
\frac{t(\epsilon-\epsilon_{\mathbf{k}\alpha})(\epsilon-\epsilon_{\mathbf{k}\alpha}+\Delta\epsilon)-2t^2}{\Lambda}
F_\alpha^{(n)}(\epsilon_{\mathbf{k}\alpha})
\]
\[\Sigma_{ii}^{f(n)}=\sum_{\mathbf{k}\alpha}\limits
|V_{i\mathbf{k}}^{\alpha}|^2
\frac{t(\epsilon-\epsilon_{\mathbf{k}\alpha})(\epsilon-\epsilon_{\mathbf{k}\alpha}-\Delta\epsilon)-2t^2}{\Lambda}
F_\alpha^{(n)}(\epsilon_{\mathbf{k}\alpha})
\]

\end{widetext}
\[\Sigma_{ii}^{d(I)}=\sum_{\mathbf{k}\alpha}\limits
\frac{|V_{i\mathbf{k}}^{\alpha}|^2}{\epsilon-\epsilon_{\mathbf{k}\alpha}}
f_\alpha(\epsilon_{\mathbf{k}\alpha})
\]
with
$\Lambda=(\epsilon-\epsilon_{\mathbf{k}\alpha})[(\epsilon-\epsilon_{\mathbf{k}\alpha}+\Delta\epsilon)
(\epsilon-\epsilon_{\mathbf{k}\alpha}-\Delta\epsilon)-4t^2]$,
$\Delta\epsilon=\epsilon_1-\epsilon_2$,
$F_\alpha^{(n)}(\epsilon_{\mathbf{k}\alpha})=f_\alpha(\epsilon_{\mathbf{k}\alpha})$
for $n=I$ and $F_\alpha^{(n)}(\epsilon_{\mathbf{k}\alpha})=1$ for
$n=0$. The self energy $\Sigma_{ij}^{(0)}$ is the self energy of
the noninteracting system, i.e., $U=0$ in the Hamiltonian (1).

Assuming that $\epsilon_1=\epsilon_2=\epsilon_0$ the self-energies
including Fermi distribution function can be calculated
analytically and expressed by means of digamma function.

Although, presented here EOM approach gives qualitatively good (physical) results around $T_K$, it breaks down (both qualitatively and quantitatively) at lower temperatures (especially at $T\ll T_K$).
We must point out that the drawbacks in EOM method are due to the logarithmic divergence of the digamma function. Detailed analysis can be found in Ref.[\onlinecite{lobo,costi,lacroix,luo}]. Here, we only list them briefly. Specifically, the divergence of the digamma function leads to wrong behavior of the linear conductance (density of states at the Fermi level)\cite{lobo}. Slave-boson calculations (which are exact at T=0K) show that the conductance saturates as the dot's level is decreased (see Fig.\ref{Fig:3}). This is no longer true for EOM approaches, where below certain dot's level position the conductance starts to decrease\cite{lobo}.
The EOM methods, with different decoupling schemes\cite{meir2,lacroix,luo}, also do not conserve completeness relation as well as not satisfy Friedel sum rule\cite{lobo2}. However, the decoupling schemes considered in our work somehow give surprisingly good dependence of the dot's occupation numbers as a function of the impurity level position (similar to those obtained by numerical renormalization group method\cite{lobo}). Taking all above into account one should remember that presented here EOM method leads to incorrect predictions at low temperatures and is limited to $T\gtrsim T_K$ as we stated in Section \ref{Sec:5}.
However, the EOM methods have an advantage on the other techniques used to investigate Kondo problem. Specifically, it enables to explore the non-equilibrium phenomena present in QDs systems biased by a finite voltage difference attached to the external leads.


\begin{thebibliography}{40}


\bibitem{cronewett} S. M. Cronenwett et al., Science {\bf281}, 540 (1998); S. Sasaki, S.
De Franceschi, J. M. Elzerman, W. G. van der Wiel, M. Eto, S.
Tarucha, and L. P. Kouvenhoven, Nature (London) {\bf 405}, 764
(2000).

\bibitem{gores} J. Gores, D. Goldhaber-Gordon, S. Heemeyer, M. A. Kastner, H.
Shtrikman, D. Mahalu, and U. Meirav, Phys. Rev. B {\bf 62}, 2188
(2000).

\bibitem{glazman} L. I. Glazman and M. E. Raikh, JETP Lett. {\bf 47},
452 (1988); T. K. Ng and P. A. Lee, Phys. Rev. Lett. {\bf 61}, 1768 (1988).

\bibitem{meir2} Y. Meir, N. S. Wingreen, and P. A. Lee, Phys. Lett. {\bf
66},3048 (1991); Phys. Lett. {\bf 70}, 2601 (1993).

\bibitem{kang} K. Kang and B. I. Min, Phys. Rev. B {\bf 52},
10689 (1995).

\bibitem{meir3} P. Nordlander, M. Pustilnik, Y. Meir, N. S. Wingreen, and
D. C. Langreth, Phys. Lett. {\bf 83}, 808 (1999).

\bibitem{aguado} R. Aguado and D. C. Langreth, Phys. Rev. Lett. {\bf
85}, 1946 (2000); R. L\'opez, R. Aguado, and G. Platero, Phys.
Rev. Lett. {\bf 89}, 136802 (2002); R. Aguado and D. C. Langreth,
Phys. Rev. B {\bf 67}, 245307 (2003).

\bibitem{aono} T. Aono and M. Eto, Phys. Rev. B {\bf 63}, 125327 (2001).


\bibitem{swirkowicz1} R. \'Swirkowicz, J. Barna\'s, and M. Wilczy\'nski ,
 Phys. Rev. B {\bf 68}, 195318 (2003); R. \'Swirkowicz, M. Wilczy\'nski, M. Wawrzyniak, and J. Barna\'s,
 Phys. Rev. B {\bf 73}, 193312 (2006); R. \'Swirkowicz, M. Wilczy\'nski, and J. Barna\'s,
 J. Phys.: Condens. Matter {\bf 18}, 2291 (2006).

\bibitem{kuzmenko} T. Kuzmenko, K. Kikoin, Y. Avishai, Phys. Rev. B {\bf 69},
195109 (2004); Phys. Rev. Lett. {\bf 96}, 046601 (2006).

\bibitem{lim} J. S. Lim, M.-S. Choi, R. L\'opez, and R. Aguado,
 Phys. Rev. B {\bf 74}, 205119 (2006).

\bibitem{zawadowski} G. Gr\"uner and A. Zawadowski, Rep. Prog. Phys. {\bf
37}, 1497 (1974); G. Zar\'ad and A. Zawadowski, Phys. Rev. Lett.
{\bf 72}, 542 (1994).

\bibitem{boese} D. Boese, W. Hofstetter, and H. Schoeller, Phys. Rev. B {\bf 64}, 125309 (2001).

\bibitem{imry} P. G. Silvestrov and Y. Imry, Phys. Rev. B {\bf 75}, 115335 (2007).

\bibitem{hubel} A.H\"ubel, K. Held, J. Weis, and K. v. Klitzing,
Phys. Rev. Lett. {\bf 101}, 186804 (2008).

\bibitem{wilhelm} U. Wilhelm, J. Schmid, J. Weis, K.v. Klitzing, Physica (Amsterdam) {\bf 14E}, 385 (2002).


\bibitem{sunG} Q.-F. Sun and H. Guo, Phys. Rev. B {\bf
66}, 155308 (2002).

\bibitem{sztenkiel} D. Sztenkiel and R. \'Swirkowicz, J. Phys.: Condens. Matter {\bf 19},
256205 (2007).

\bibitem{sztenkiel2} D. Sztenkiel and R. \'Swirkowicz, J. Phys.: Condens. Matter {\bf 19}, 386224 (2007).

\bibitem{holle} A. W. Holleitner, A. Chudnovskiy, D. Pfannkuche, K. Eberl, and R. H.
Blick, Phys. Rev. B {\bf 70}, 075204 (2004).

\bibitem{krychowski} S. Lipi\'nski and D. Krychowski, Phys. Status Solidi b {\bf
243}, 206 (2005).

\bibitem{shon} T. Pohjola, H. Schoeller, and G. Sch\"on, Europhys. Lett., {\bf 54},
241 (2001).

\bibitem{wen}
J. Wen, J. Peng, B. Wang, and D. Y. Xing, Phys. Rev. B {\bf 75},
155327 (2007).

\bibitem{kubo}
T. Kubo, Y. Tokura, and S. Tarucha, Phys. Rev. B {\bf 77},
041305(R) (2008).

\bibitem{meir92} Y. Meir, N. S. Wingreen,  Phys. Rev. Lett.
{\bf 68}, 2512 (1992).


\bibitem{hewson} A. C. Hewson, The Kondo Problem to Heavy Fermions
(Cambridge University Press, Cambridge, U.K., 1993).

\bibitem{martinek} J. Martinek, Y. Utsumi, H. Imamura, J. Barna\'s, S. Maekawa, J. K\"onig, and G.
Sch\"on, Phys. Lett. {\bf 91},127203 (2003); D. Matsubayashi and M. Eto, Phys. Rev B {\bf 75}, 165319 (2007).

\bibitem{wohlmann} V. Kashcheyevs, A. Schiller, A. Aharony, and O.
Entin-Wohlman, Phys. Rev. B {\bf 75}, 115313 (2007).

\bibitem{coleman} P. Coleman, Phys. Rev. B {\bf 29}, 3036 (1984).

\bibitem{Ng} T. K. Ng, Phys. Rev. Lett. {\bf }, 3635 (1993).

\bibitem{trocha}
P. Trocha, J. Barna\'s, Phys. Rev. B {\bf 76}, 165432 (2007).

\bibitem{trochaJNN}
P. Trocha and J. Barna\'s, J. Nanosci. Nanotechnol. {\bf 10}, 2489 (2010).

\bibitem{lacroix} C. Lacroix, J. Appl. Phys {\bf 53}, 2131 (1982).

\bibitem{costi} T. A. Costi, J. Phys. C: Solid State Phys. {\bf 19}, 5665 (1986).

\bibitem{lobo} T. Lobo, M.S. Figueira, R. Franco, J. Silva-Valencia, and M.E. Foglio, Physica B {\bf 398}, 446 (2007).

\bibitem{luo} H.-G. Luo, J.-J. Ying, and S.-J. Wang, Phys. Rev. B {\bf 59}, 9710 (1999).

\bibitem{lobo2} T. Lobo, M. S. Figueira, and M. E. Foglio, Nanotechnology {\bf 17}, 6016 (2006); \emph{ibib} {\bf 21}, 274007 (2010).

\end{thebibliography}
\end{document}